\documentclass[twoside,reqno]{bjp}
\usepackage{epsfig,cite}
\usepackage{graphics}
\usepackage{amssymb,amsmath}
\usepackage{times}
\begin{document}

\sloppy \raggedbottom

 \setcounter{page}{1}

\title{Proxy-SU(3) symmetry in heavy nuclei: Prolate dominance and prolate-oblate shape transition}

\runningheads{Prolate dominance and prolate-oblate shape transition}{S. Sarantopoulou, D. Bonatsos, I.E. Assimakis, N. Minkov, A. Martinou, et al.}

\begin{start}

\author{S. Sarantopoulou}{1},
\coauthor{D. Bonatsos}{1},
\coauthor{I.E. Assimakis}{1},
\coauthor{N. Minkov}{2},
\coauthor{A. Martinou}{1},
\coauthor{R.B. Cakirli}{3},
\coauthor{R.F. Casten}{4,5},
\coauthor{K. Blaum}{6}

\address{Institute of Nuclear and Particle Physics, National Centre for Scientific Research ``Demokritos'', GR-15310 Aghia Paraskevi, Attiki, Greece}{1}

\address{Institute of Nuclear Research and Nuclear Energy, Bulgarian Academy of Sciences, 72 Tzarigrad Road, 1784 Sofia, Bulgaria}{2}

\address{Department of Physics, University of Istanbul, 34134 Istanbul, Turkey}{3} 

\address{Wright Laboratory, Yale University, New Haven, Connecticut 06520, USA}{4}

\address{Facility for Rare Isotope Beams, 640 South Shaw Lane, Michigan State University, East Lansing, MI 48824 USA}{5}

\address{Max-Planck-Institut f\"{u}r Kernphysik, Saupfercheckweg 1, D-69117 Heidelberg, Germany}{6}

\received{31 October 2017}

\begin{Abstract}

Using a new approximate analytic parameter-free proxy-SU(3) scheme, simple predictions  for the global feature of prolate dominance and for the locus of the prolate-oblate shape transition have been made and compared with empirical data. Emphasis is placed on the mechanism leading to the breaking of the particle-hole symmetry, 
which is instrumental in shaping up these predictions. It turns out that this mechanism is based on the SU(3) symmetry and the Pauli principle alone, without reference to any specific Hamiltonian.

\end{Abstract}

\PACS {21.60.Fw, 21.60.Ev, 21.60.Cs}

\end{start}
\section{Intoduction}

The recent introduction of the proxy-SU(3) scheme \cite{proxy1} has led to parameter independent predictions of the deformation parameters $\beta$ and $\gamma$ \cite{proxy2}, as well as to predictions of the locus of the prolate to oblate shape transition on the nuclear chart and the prolate over oblate dominance in deformed rare earth nuclei \cite{proxy2}. 

In Section 2 of the present work we extend the study of the prolate to oblate transition to the actinides and superheavy elements, while in Section 3 we study in detail the particle-hole symmetry breaking, which leads to the prolate over oblate dominance and to the determination of the border of the prolate to oblate transition.  

\section{Prolate to oblate transition in actinides and superheavy elements}

In order to determine in proxy-SU(3) the SU(3) irreducible representation (irrep)
corresponding to a given nucleus, one needs the highest weight (h.w.) irreps of the relevant U(n) algebras for protons and neutrons, given in Table I. This is an extension of Table~{I} of Ref. \cite{proxy2}, to which the U(28) and U(36) results have been added. These results can be obtained through the UNTOU3 code \cite{code}. For the h.w. irreps, in particular, an analytic formula exists \cite{Kota1,Kota2,Kota3}. The extra columns in  Table I can be used for the determination of the h.w. irreps in the actinides, as well as in superheavy elements (SHE) \cite{Ring,Skalski}. For the illustrative and pedagogical purposes of this work, including the text and Table II, we take the relevant shells for the actinides and super heavy nuclei as $Z = 82$-126 and $N = 126$-184 although the upper bounds are by no means certain and microscopic calculations give many varying scenarios.   

Consider $^{260}_{100}$Fm$_{160}$ as an example. This nucleus possesses $100-82=18$ valence protons in the 82-126 shell, which is approximated by the proxy-SU(3) pfh shell with U(21) symmetry. Therefore the h.w. irrep for the valence protons is (36,6). 
It also has 160-126=34 valence neutrons in the 126-184 shell, approximated by the sdgi proxy-SU(3) shell with U(28) symmetry.  Therefore the h.w. irrep for the valence neutrons is (46,16). The whole nucleus is then described by the stretched \cite{DW2} irrep (82,22). 

Results for nuclei with protons in the 82-126 shell and neutrons in the 126-184 shell are summarized in Table II. Prolate nuclei are characterized by $\lambda>\mu$, while oblate nuclei have $\lambda<\mu$ \cite{proxy2}. In Table II we see that a small region 
of oblate nuclei (underlined) appears in the lower right corner of the table, i.e., 
just below the proton shell closure and the neutron shell closure. This result is similar to what is seen in Tables II and III of Ref. \cite{proxy2} for rare earth nuclei. Beta and gamma deformation parameters for the nuclei appearing in Table II
can be found in another paper in this Workshop \cite{BonatsosSDANCA17}.

\section{Breaking of the particle-hole symmetry}\label{ph} 

The breaking of the particle-hole (p-h) symmetry plays a key role in the determination 
of the place of the prolate to oblate transition within a given shell, as seen in Ref.
\cite{proxy2} (see especially Fig. 1 in this reference). This topic has been briefly discussed in Ref. \cite{EPJA}, but it does deserve a more detailed discussion.

In a simple way, one can say that the highest weight (h.w.) irrep corresponds to the most probable distribution of a given number of particles over the available levels in a given harmonic oscillator (h.o.) shell. This is explained in a clear way in Fig. 1  of Ref. \cite{code}.

In more detail, let us consider the $n$-th shell of the harmonic oscillator, which possesses and overall 
U((n+1)(n+2)/2) symmetry. For example, the sd shell has $n=2$ and possesses the U(6) symmetry, the pf shell has $n=3$ and possesses the U(10) symmetry, and so on.  

Let us consider the U(6) case in more detail. 
If we have a system of bosons with U(6) symmetry, as in the Interacting Boson Model (IBM) \cite{IA}, one boson will give the (2,0) h.w. irrep, 2 bosons will give (4,0), 3 bosons will give (6,0), 4 bosons will give (8,0), 
N bosons will give (2N,0), as in IBM.
This happens because bosons will crowd the most symmetric irrep of U(6), characterized by a Young diagram with only one line (fully symmetric). For 1, 2, 3, 4 bosons the U(6) irreps are [1], [2], [3], [4] respectively.   

If we now consider a system of (like) fermions with U(6) symmetry, one fermion will give (2,0), 2 fermions will give (4,0), 3 fermions will give (4,1), 4 fermions will give (4,2), as in Table I. This is a consequence of the Pauli principle. 
Because of antisymmetrization, the U(6) irreps in this case can have only 2 columns, 
thus for 1, 2, 3, 4 fermions the U(6) irreps will be [1], [2], [21], [22], as in 
Table I. 

We see that the differences between the boson irreps and the fermion irreps, both in U(6)
and in its SU(3) subalgebra, arise solely from the Pauli principle, without reference to any specific Hamiltonian.  

In a similar way, the particle-hole asymmetry arises because of the restrictions imposed 
by the Pauli principle, irrespectively of any specific Hamiltonian. 

Let us now consider a pf shell and start filling it with fermions. One fermion will have the (3,0) h.w. irrep, 2 fermions will have the (6,0) irrep, 3 fermions will have the (7,1) irrep, 4 fermions will have the (8,2) irrep, 5 fermions will have the (10,1) irrep, 
6 fermions will have the (12,0) irrep, as in Table 1, the corresponding U(10) irreps being [1], [2], [21], [22], [221], [222] respectively.
By the way, if we had bosons, the U(10) irreps would have been [1], [2], [3], [4], [5], [6], and the SU(3) irreps (3,0), (6,0), (9,0), (12,0), (15,0), (18,0) respectively.
The consequences of the Pauli principle are again vividly present.

We go on gradually filling the pf shell and we end up with the results of Table I. These indicate that up to 4 particles or 4 holes, there is a particle-hole 
symmetry, but the symmetry is broken beyond this point. Actually this p-h symmetry 
is seen up to 4 particles or 4 holes in all U(n) symmetries of the h.o., irrespectively of n.  

Full results of the irreps occurring for 2, 4, 6, 8, 12, 14, 16, 18 particles are given in Table III, produced by the code of Ref. \cite{code}. Several comments apply. 

1) In the case of 2 and 18 particles, 2 irreps appear in each case. The h.w. irreps 
are conjugate to each other, as it should be expected with p-h symmetry present.

2) In the case of 4 and 16 particles, 11 different irreps appear (some of them more than once, as indicated by the exponents, which stand for multiplicity numbers).  The irreps appearing for 16 particles are the conjugates of the irreps appearing for 4 particles, but they do not appear in the same order. However, the h.w. irrep (2,8)  
in the case of 16 particles is the conjugate of the (8,2) irrep, which is the h.w. irrep
in the case of 4 particles. Thus the p-h symmetry is preserved, but only as far as the h.w.
irrep is concerned. 

3) In the case of 6 and 14 particles, 24 different irreps appear (some of them more than once, as indicated by the exponents in Table 3, which indicate multiplicities of irreps).  The irreps appearing for 14 particles are the conjugates of the irreps appearing for 6 particles, but they do not appear in the same order. Furthermore, while the (12,0)
irrep is the h.w. irrep for 6 particles, its conjugate, (0,12), appears in the 8th place 
in the case of 14 particles. In addition, the (6,6) irrep, which is the h.w. irrep for 14 particles, appears in the 3rd place in the case of 6 particles. In other words, while the (12,0) irrep is the most probable one for 6 particles, for 14 particles the (6,6) irrep
is the most probable one. This is a consequence of the Pauli principle alone, irrespectively of any Hamiltonian. The restrictions imposed by the Pauli principle in the lower half of the shell, where the shell is relatively empty and the particles have more choices at their disposal, leads to a result different from the one in the upper half of the shell, where particles get crowded and have fewer choices at their disposal. 

4) The same asymmetry is seen in the case of 8 and 12 particles. The sets of 32 different irreps appearing in these two cases are still conjugate to each other, but the order in which they appear and the h.w. irrep are different for each particle number. 

Similar conclusions can be drawn from the lists of SU(3) irreps occurring in the sdg shell with U(15) symmetry, given in Table IV.

From the above it becomes clear that a particle-hole symmetry holds as far as the set of irreps appearing in each case is concerned. However, the p-h symmetry is broken as far as the order of appearance of the irreps according to their weight is concerned. As a result,
the h.w. irrep is different if more than 4 particles or holes are present. This is a result 
imposed by the Pauli principle alone, without involving any specific Hamiltonian. 

In earlier work, the argument was used that the irrep with the highest eigenvalue of the second order Casimir operator of SU(3) should be used, because it corresponds to the highest value of the quadrupole-quadrupole interaction, through the well-known relation \cite{Draayer}
\begin{equation}
C_2= {1\over 4} Q\cdot Q + {3\over 4} L^2,
\end{equation}
where $C_2$ stands for the second order Casimir operator of SU(3), $Q$ is the quadrupole operator and $L$ denotes the angular momentum. 
This choice gives identical results with the h.w. choice up to the middle of the shell,
as one can see in Table I, in which the irreps corresponding to the highest eigenvalue 
of $C_2$ are given in the columns labelled by $C$.  
Beyond the middle of the shell the $C_2$ choice gives results corresponding to the p-h symmetry,
since obviously the irrep with the highest $C_2$ eigenvalue will be the same one 
both in the upper and in the lower part of the shell, since the expression for the eigenvalues is symmetric in $\lambda$ and $\mu$, namely \cite{Draayer}
\begin{equation}
C_2(\lambda,\mu) = (\lambda+\mu+3)(\lambda +\mu) -\lambda \mu.
\end{equation}
As an example, consider the case of 8 and 12 particles in the pf shell. For 8 particles, the h.w. irrep is the (10,4) one. One can easily see that this is the irrep with the highest $C_2$ eigenvalue
among all the irreps appearing for 8 particles. In the upper part of the shell, the irrep 
(4,10) still has the highest $C_2$ eigenvalue, but the Pauli principle has pushed it down 
to 4th place as far as the weight, i.e. the probability of appearance, is concerned. The 
h.w. irrep in this case is (12,0), despite the fact that it possesses a lower $C_2$ eigenvalue. 

From the above it becomes clear that the p-h symmetry breaking appearing in the proxy approach is imposed by the Pauli principle alone, without reference to any particular Hamiltonian. In contrast, the choice of the irrep with the highest $C_2$ eigenvalue is based on a particular 
choice of the Hamiltonian. Fortunately, most of the applications in earlier work have been carried out in the lower half of various shells, thus the bulk of the relevant results remains perfectly valid. 

It is well known \cite{code} 
that the h.w. irrep has to be a unique one, i.e., it has to occur only once. From Table
3  we see that the percentage of different irreps which qualify as candidates for the h.w. place is reduced with increasing number of particles within the lower half of the shell. For 2 particles there are 2 unique irreps out of 2 (100\%), 
for 4 particles there are 9 unique irreps out of 11 (81.2\%), for 6 particles there are 
9 unique irreps out of 24 (37.5\%), for 8 particles there are 8 unique irreps out of 32 
(25\%). Once more, similar conclusions can be drawn from the lists of SU(3) irreps occurring in the sdg shell with U(15) symmetry, given in Table IV.

\section{Conclusions}

The particle-hole symmetry breaking, which is instrumental in causing the prolate over oblate dominance in deformed nuclei and in defining the locus of the prolate to oblate shape transition, is found to be a consequence of the SU(3) symmetry and the Pauli principle, without reference to any specific Hamiltonian. 

\section*{Acknowledgements}

Work partly supported by the Bulgarian National Science Fund (BNSF) under Contract No. DFNI-E02/6, by the US DOE under Grant No. DE-FG02- 91ER-40609, and by the MSU-FRIB laboratory, by the Max Planck Partner group, TUBA-GEBIP, and by the Istanbul University Scientific Research Project No. 54135.


\begin{figure}[t]
{\epsfig{file=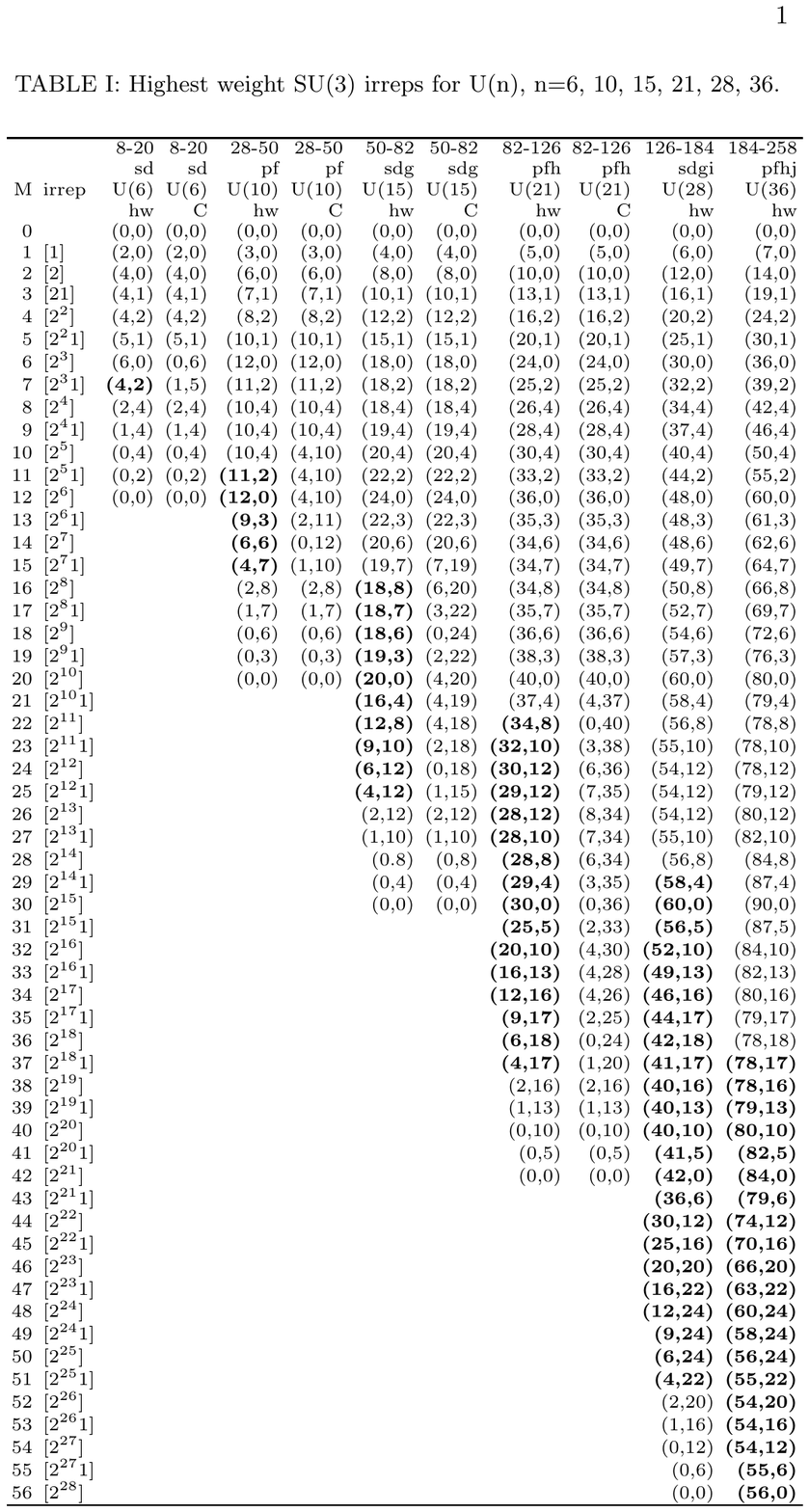,width=180mm}}

\end{figure}


\begin{figure}[t]
\centering{\epsfig{file=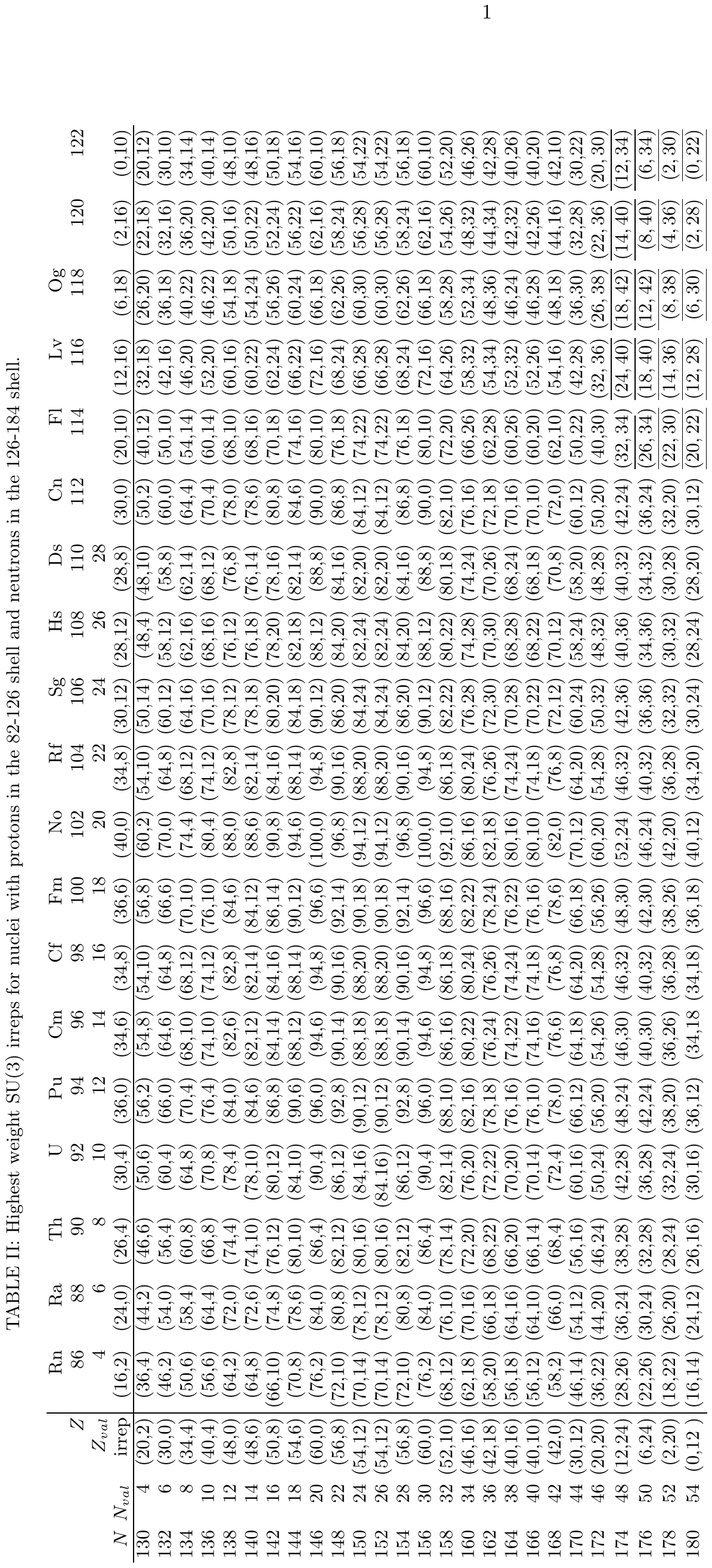,width=140mm}}

\end{figure}


\begin{figure}[t]
\centering{\epsfig{file=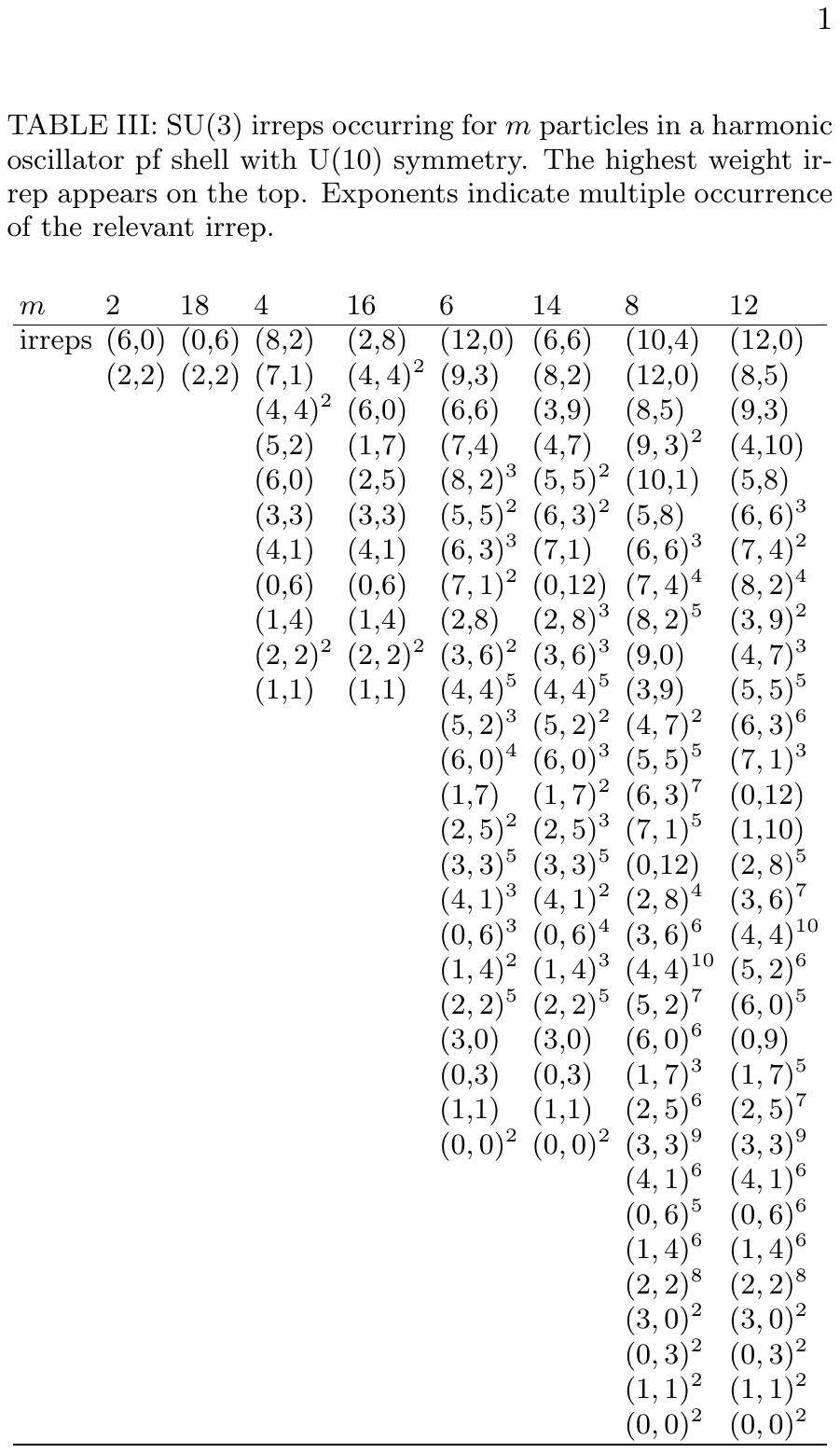,width=155mm}}

\end{figure}


\begin{figure}[t]
\centering{\epsfig{file=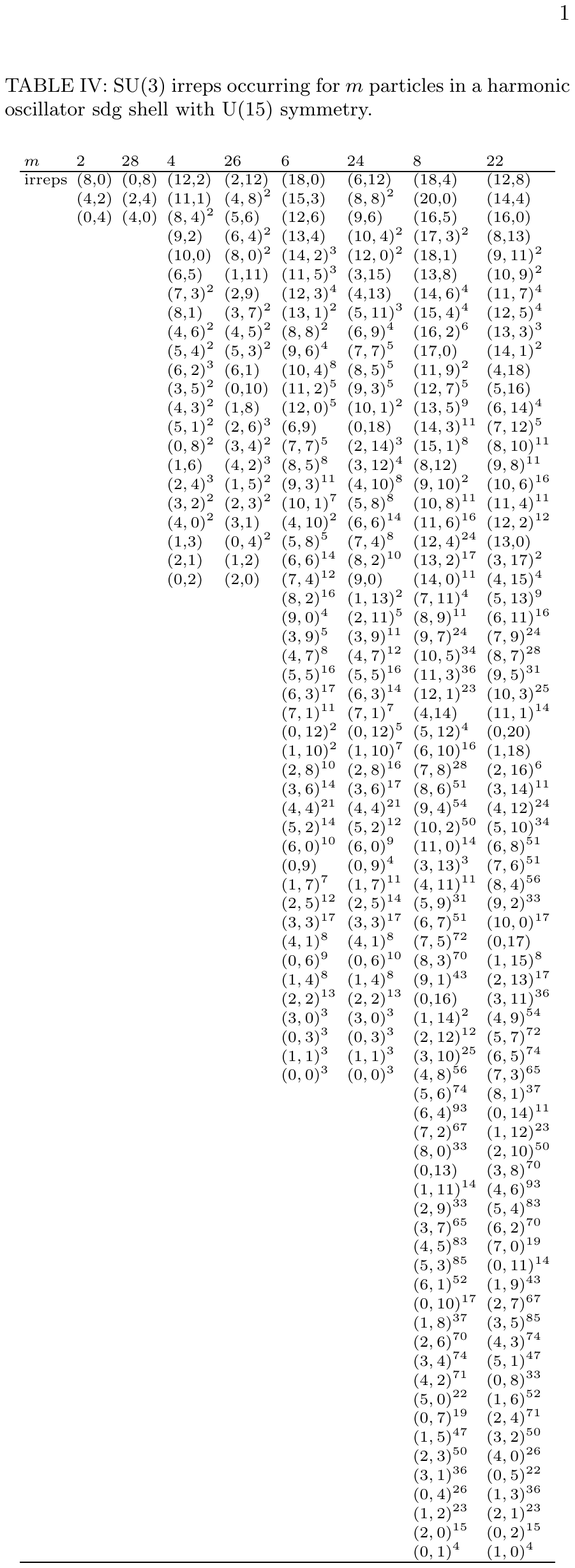,width=155mm}}

\end{figure}

\end{document}